# Phase-controlled atom-photon entanglement in a three-level V-type atomic system via spontaneously generated coherence

**Mohammad Abazari[1], Ali Mortezapour [2], Mohammad Mahmoudi [1],* and Mostafa Sahrai[3]**

1. Physics Department, University of Zanjan, P. O. Box 45195-313, Zanjan, Iran
2. Institutes for Advanced Studies in Basic Sciences, P.O. Box 45195-159, Zanjan, Iran
3. Research Institute for Applied Physics and Astronomy, University of Tabriz, Tabriz, Iran

E-Mail: a.mortezapour7@gmail.com
E-Mail: mahmoudi@znu.ac.ir

**Abstract:** We investigate the dynamical behavior of the atom-photon entanglement in a V-type three-level quantum system using the atomic reduced entropy. It is shown that an atom and photons are entangled at the steady-state; however disentanglement can also be achieved in an especial condition. It is demonstrated that in the presence of quantum interference induced by spontaneous emission, the reduced entropy and the atom-photon entanglement are phase-dependent. Non-stationary solution is also obtained when the quantum interference due to the spontaneous emission is completely included.

**Keywords:** Quantum interference, Quantum entanglement, Quantum entropy, Phase- dependent.

**PACS Codes:** 03.67.Mn, 42.50.Dv, 42.50.Gy

## 1. Introduction

Quantum correlation between different parts of a system leads to an important quantum phenomenon known as entanglement. Entanglement allows having a much closer relationship than is possible in classical physics. A system consisting of two components is said to be entangled if its quantum state cannot be described by a simple product of the quantum states of the two components [1]. Under this circumstance measurement on the one of them gives information about other component. For a bicomponent system in a pure state, it has been shown that the reduced quantum entropy is the best tool for measure the degree of entanglement between two components [2]. The higher reduced quantum entropy means the higher degree of entanglement.

In bicomponent systems the entanglement can be established between two particles or between the particle and the field. The Einstein-Podolsky-Rosen (EPR) state [3] is an interesting example of two- particle components entanglement which can be used in secure quantum communication prescript [4]. Note that quantum entanglement can also be generated in a system with three or more components [5]. Quantum entanglement is the basic concept of the quantum information processes, such as quantum computing [6], quantum teleportation [7], quantum cryptography [8] and quantum communication [9]. The atom-photon entanglement has been studied in atomic cascade systems [10, 11] as well as in trapped ions [12-13]. The observation of the quantum entanglement between a single trapped $^{87}Rb$ atom and a single photon at a wavelength suitable for low-loss communication has been reported [14].

Theoretical description of entanglement evolution between atom and quantized field in the Jaynes-Cummings model has been proposed [15-17]. However, it was shown that the induced entanglement between two interacting two-level quantum systems can be controlled by the relative phase of applied fields [18]. In another study it was shown that the atom-photon entanglement near a 3D anisotropic photonic band edge depends on the relative phase of applied fields [19].

Generally, the entanglement can be controlled by the initial condition of the atomic states. Atomic coherence and quantum interference are the basic mechanisms for controlling the optical properties of the medium. In fact, the discovery of electromagnetically induced transparency has opened up a new rote to control the optical properties of atom-photon coherent interaction [20, 21]. Quantum interference induced by spontaneous emission, however, can modify the response of atom-photon entanglement. Vacuum induced coherence, i. e. spontaneously generated coherence (SGC), [22] can also make the system phase dependent [23]. Various schemes have been proposed for phase control of optical properties such as like light propagation [24], transient behavior of the medium [25, 26], probe gain [27, 28], and phase dependent of resonance fluorescence spectrum [29].

Recently, two of the present authors, with collaborators, have investigated the dynamical behavior of the dispersion and the absorption in a V-type three-level atomic system in the presence of quantum interference induced by the spontaneous emission. It was shown that in the presence of decay-induced interference the probe dispersion and absorption are completely phase-dependent [30].

The effect of quantum interference on the entanglement of a driven V-type three-level atom and its spontaneous emission field were investigated by using the concept of quantum entropy [31]. He has shown that for appropriate atomic parameters they can be entangled or disentangled at the steady-state. In this paper, we investigate the phase-dependent atom-photon entanglement in a V-type three-level atomic system in the presence of the quantum interference due to the spontaneous emission. It is demonstrated that such entanglement can be controlled just by changing the relative phase of applied fields. In addition, we found that under the special condition the atom and photons become disentangled.

## 2. Model and Solution

Consider a three-level V-type atomic system (Figure 1(a)) with a ground state $|1\rangle$, and two excited-states $|2\rangle$, $|3\rangle$. The quantum system is coupled by two classical fields. The left field $\vec{E}_L = E_L \hat{e}_L e^{-i(\omega_R t - \vec{K}_R \cdot \vec{r} + \varphi_L)} + c.c$ with frequency $\omega_L$ and Rabi frequency $\Omega_L$ drives the transition $|1\rangle \to |3\rangle$, and other right field, $\vec{E}_R = E_R \hat{e}_R e^{-i(\omega_R t - \vec{K}_R \cdot \vec{r} + \varphi_R)} + c.c$ with frequency $\omega_R$ and Rabi frequency $\Omega_R$ is applied to the transition $|1\rangle \to |2\rangle$. Here, $E_L(E_R)$ and $\hat{e}_L(\hat{e}_R)$ are the amplitude, and the polarization of the left (right) classical laser field, while $\omega_L(\omega_R)$, $\vec{k}_L(\vec{k}_R)$ and $\varphi_L(\varphi_R)$ are the frequency, wave vector, and initial phase of left (right) classical laser field. The parameters $2\gamma_{21}$ and $2\gamma_{31}$ denote the spontaneous decay rates from excited-states $|2\rangle$ and $|3\rangle$ to ground state $|1\rangle$, respectively. Also $\Delta_L = \omega_L - \omega_{31}$, $\Delta_R = \omega_R - \omega_{21}$ are one-photon detuning of the two fields. Such a system, with a single ground state and a closely spaced excited doublet (e.g. two near-degenerate states), is damped by the usual vacuum interactions, so the two decay pathways from the excited doublet to the ground state are not independent. The system decays from the upper states doublet to a lower state via spontaneous emission leading to the quantum interference, i.e. spontaneously generated coherence (SGC) [22].

In the following, we consider the sodium $D_2$ transition as a realistic example. The decay rate of transition is $\gamma = 2\pi \times 9.79 MHz$. The right field $\Omega_R$ is applied to the $3\ ^2S_{1/2} - 3\ ^2P_{1/2}$ transition, while the left field $\Omega_L$ is applied to the $3\ ^2S_{1/2} - 3\ ^2P_{3/2}$ transition. For such a transition we have $\omega_{32} = 0.2\gamma \cong 15.8 MHz$. Note that the two upper levels are near-degenerate, so the quantum interference due to the spontaneous emission can be induced [30]. The interaction Hamiltonian describing the dynamics of the system in the dipole and rotating-wave approximations and rotating frame is given by:

$$H = \hbar\{\omega_1|1\rangle\langle 1| + (\omega_1 + \omega_R - \Delta_R)|2\rangle\langle 2| + (\omega_1 + \omega_L - \Delta_L)|3\rangle\langle 3| \\ -\Omega_R^* e^{i\varphi_R}|1\rangle\langle 2| - \Omega_L^* e^{i\varphi_L}|1\rangle\langle 3|\} + H.c. \tag{1}$$

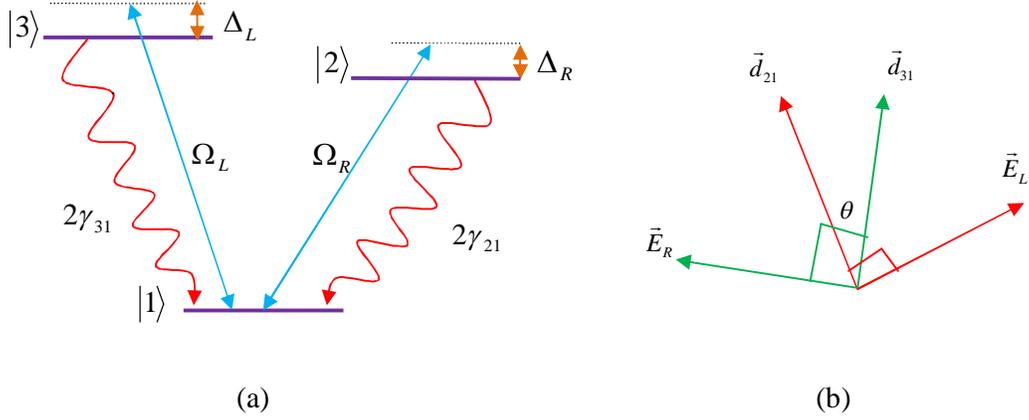

**Figure 1.** A V-type three-level atom driven by two laser fields with corresponding Rabi frequencies $\Omega_R$, $\Omega_L$. (b) The arrangement of field polarization for a single field driving one transition if dipoles are nonorthogonal.

The density matrix equations of motion for the atomic variables can be written as:

$$\dot{\rho}_{22}(t) = -2\gamma_{21}\rho_{22}(t) + i\{\Omega_R \rho_{12}(t) - \Omega_R^* \rho_{21}(t)\} - K_c\sqrt{\gamma_{21}\gamma_{31}}(\rho_{23}(t)e^{-i\varphi} + \rho_{32}(t)e^{i\varphi}),$$

$$\dot{\rho}_{33}(t) = -2\gamma_{31}\rho_{33}(t) + i\{\Omega_L \rho_{13}(t) - \Omega_L^* \rho_{31}(t)\} - K_c\sqrt{\gamma_{21}\gamma_{31}}(\rho_{23}(t)e^{-i\varphi} + \rho_{32}(t)e^{i\varphi}),$$

$$\dot{\rho}_{12}(t) = -(\gamma_{21} + i\Delta_R)\rho_{12}(t) + i\Omega_R^*(\rho_{22}(t) - \rho_{11}(t)) + i\Omega_L^* \rho_{32}(t) - K_c\sqrt{\gamma_{21}\gamma_{31}}\rho_{13}(t)e^{-i\varphi},$$

$$\dot{\rho}_{13}(t) = -(\gamma_{31} - i(\delta - \Delta_L))\rho_{13}(t) + i\Omega_L^*(\rho_{33}(t) - \rho_{11}(t)) + i\Omega_R^* \rho_{23}(t) - K_c\sqrt{\gamma_{21}\gamma_{31}}\rho_{12}(t)e^{i\varphi},$$

$$\dot{\rho}_{32}(t) = -(\gamma_{21} + \gamma_{31} + i(\Delta_R - \Delta_L + \delta))\rho_{32}(t) + i\Omega_L \rho_{12}(t) - i\Omega_R^* \rho_{31}(t) - K_c\sqrt{\gamma_{21}\gamma_{31}}(\rho_{22}(t) + \rho_{33}(t))e^{-i\varphi},$$

$$\rho_{11}(t) + \rho_{22}(t) + \rho_{33}(t) = 1,$$
(2)

where $\varphi = \varphi_R - \varphi_L$ and $\delta = \omega_L - \omega_L$ are the relative phase and the relative frequency of the driving fields, respectively.

The strength of the quantum interference resulting from the cross coupling between the transitions $|2\rangle \to |1\rangle$ and $|3\rangle \to |1\rangle$ is measured by the parameter $K_c = \vec{d}_{21}\cdot\vec{d}_{31}/|\vec{d}_{21}||\vec{d}_{31}| = \cos\theta$, where $\vec{d}_{21}$ and $\vec{d}_{31}$ are the dipole moments of the corresponding transitions and $\theta$ is the angle between the two induced dipole moments as shown by Figure 1(b). The effects of quantum interference are sensitive to the orientations of the atomic dipole moments. For parallel dipole moments, the interference effect is maximal and $K_c = 1$, while for perpendicular dipole moments, $K_c = 0$, and the quantum interference disappears. Note that the relative phase appears through equations via the parameter $K_c$. So,

in Eqs (2), the effect of relative phase of applied fields appear in all terms contain $K_c$. Then the solutions of these equations for $K_c \neq 0$ are phase-dependent.

Now, we seek the corresponding steady state analytical solution for elements of density matrix for the weak Rabi-frequencies and $\Omega_R = \Omega_L = \Omega_0$. The population and coherence terms of density matrix for $\gamma = \gamma_{21} = \gamma_{31} = 1.0$, $\Delta_R = \Delta_L = 0$ are given by:

$$\rho_{22} = \rho_{33} = \frac{\Omega_0^2(K_c(2+\Omega_0^2)\cos\varphi - \Omega_0^2 - (1+K_c^2))}{K_c\Omega_0^2(7+4\Omega_0^2+K_c^2)\cos\varphi - K_c^2(K_c^2-2+3\Omega_0^2) - \Omega_0^2 - (1+2\Omega_0^2)^2},$$

$$\rho_{12} = \rho_{13} = \frac{-i\Omega_0(K_c(2\Omega_0^2 K_c \cos\varphi - (\Omega_0^2 + 2(K_c^2-1)))e^{-i\varphi} + K_c\Omega_0^2 e^{i\varphi} + 2((K_c^2-1)-\Omega_0^2))}{2K_c\Omega_0^2(7+4\Omega_0^2+K_c^2)\cos\varphi - 2K_c^2(K_c^2-2+3\Omega_0^2) - 2\Omega_0^2 - 2(1+2\Omega_0^2)^2},$$

$$\rho_{23} = \frac{\Omega_0^2(\Omega_0^2 K_c \cos\varphi - K_c e^{i\varphi}(K_c e^{i\varphi}-2) - (1+\Omega_0^2))}{K_c\Omega_0^2(7+4\Omega_0^2+K_c^2)\cos\varphi - K_c^2(K_c^2-2+3\Omega_0^2) - \Omega_0^2 - (1+2\Omega_0^2)^2},$$

$$\rho_{11} + \rho_{22} + \rho_{33} = 1. \tag{3}$$

All expressions in Eqs. (3) are defiantly phase-dependent.

## 3. Entropy and Entanglement

A bicomponent system is described by a density matrix of a $(\mathbf{C}^m \otimes \mathbf{C}^n)$ Hilbert space. The partial density matrix of one part is obtained by tracing over other [32]:

$$\rho_{A(B)} = Tr_{B(A)}(\rho_{AB}). \tag{4}$$

A bipartite quantum system is considered separable, if it can be written as:

$$\rho_{AB} = \rho_A \otimes \rho_B, \tag{5}$$

where $\rho_{A(B)}$ are the individual partial density matrixes. If the system can not satisfy Eq. (5), it is said to be entangled. The atom-field quantum entanglement can be discussed by using Von Neumann entropy which is defined as [33]:

$$S = -Tr(\rho \ln \rho), \tag{6}$$

where $\rho$ is the density matrix operator. For a pure state the entropy is vanished while for a mixture state is nonzero. We assume that the quantum entropy of total system is zero corresponding to a pure state, while the partial entropy of subsystem varies with time. According to the triangle inequality [34]:

$$|S_A(t) - S_F(t)| \leq S_{AF}(t) \leq |S_A(t) + S_F(t)|, \tag{7}$$

For a closed system that starts in a pure state, partial entropies of the field and atom are equal at all times after beginning of interaction between two subsystems. Then our information about the entropy of each subsystems leads to the entanglement between the subsystems. Phoenix and Knight [35-36] have shown that, under these circumstances a decrease in partial entropy means that each subsystem evolves toward a pure quantum state, whereas a rise in partial entropy means that the two components tend to lose their individuality and become correlated or entangled. The degree of entanglement (DEM) for atom-field entanglement is defined as:

$$DEM(t) = S_A = S_F = -\sum_{j=1}^{3} \lambda_j \ln \lambda_j. \qquad (8)$$

## 4. Results and Conclusion

We now summarize our results for the steady state behavior of the system in Eqs. (1)-(5). For simplicity, all parameters are reduced to dimensionless units through scaling by $\gamma_{21} = \gamma_{31} = \gamma$ and all figures are plotted in the unit of $\gamma$. We assume the applied fields have a same frequency.

We first investigate the effect of quantum interference due to the spontaneous emission on phase control of the quantum entropy. In Figure 2, we display the time dependent behavior of the quantum entropy for different relative phase of applied fields. The using parameters are $\Omega_R = \Omega_L = 0.1\gamma$, $\delta = 0.0$, $\Delta_R = \Delta_L = 0.0\gamma$ (left column), $\Delta_R = \Delta_L = 2.0\gamma$ (right column), $\varphi = 0$ (solid), $\varphi = \pi/6$ (dashed), $\varphi = 4\pi/3$ (dotted) and (a, d) $K_c = 0$, (b, e) $K_c = 0.5$, (c, f) $K_c = 0.99$. An investigation on Figure 2 shows that in the absence of quantum interference due to the spontaneous emission, the quantum entropy is phase-independent, while by including the effect of quantum interference, the entropy changes by the changing of relative phase of applied fields. Moreover, in Figures 2(c, f) for $\varphi = 0$ the quantum entropy vanishes which corresponds to the disentanglement. By comparing Figures 2 (a)-(c) for different values of $K_c$, we observe that the quantum interference between two spontaneous emissions has a major role in establishing the atom-photon entanglement.

On the other hand, comparing the left and right columns in Figure 2 shows that by increasing the detuning of external applied field, the entropy decreases. To investigate the effect of one-photon transition on atom–photon entanglement, we plot the steady state entropy versus detuning of applied fields for different values of $K_c$ in Figure 3. The selected parameters are same as in Figure 2. The maximum entanglement is occurred in one-photon resonance condition.

**Figure 2.** Time dependent behavior of the quantum entropy for different relative phase of applied fields. The selected parameters are $\gamma = 1, \gamma_{21} = \gamma_{31} = \gamma$, $\Omega_R = \Omega_L = 0.1\gamma$, $\delta = 0.0\gamma$, $\Delta_R = \Delta_L = 0.0\gamma$ (left column), $\Delta_R = \Delta_L = 2.0\gamma$ (right column), $\varphi = 0$ (solid), $\varphi = \pi/6$ (dashed), $\varphi = 4\pi/3$ (dotted), (a, d) $K_c = 0$, (b, e) $K_c = 0.5$, and (c, f) $K_c = 0.99$.

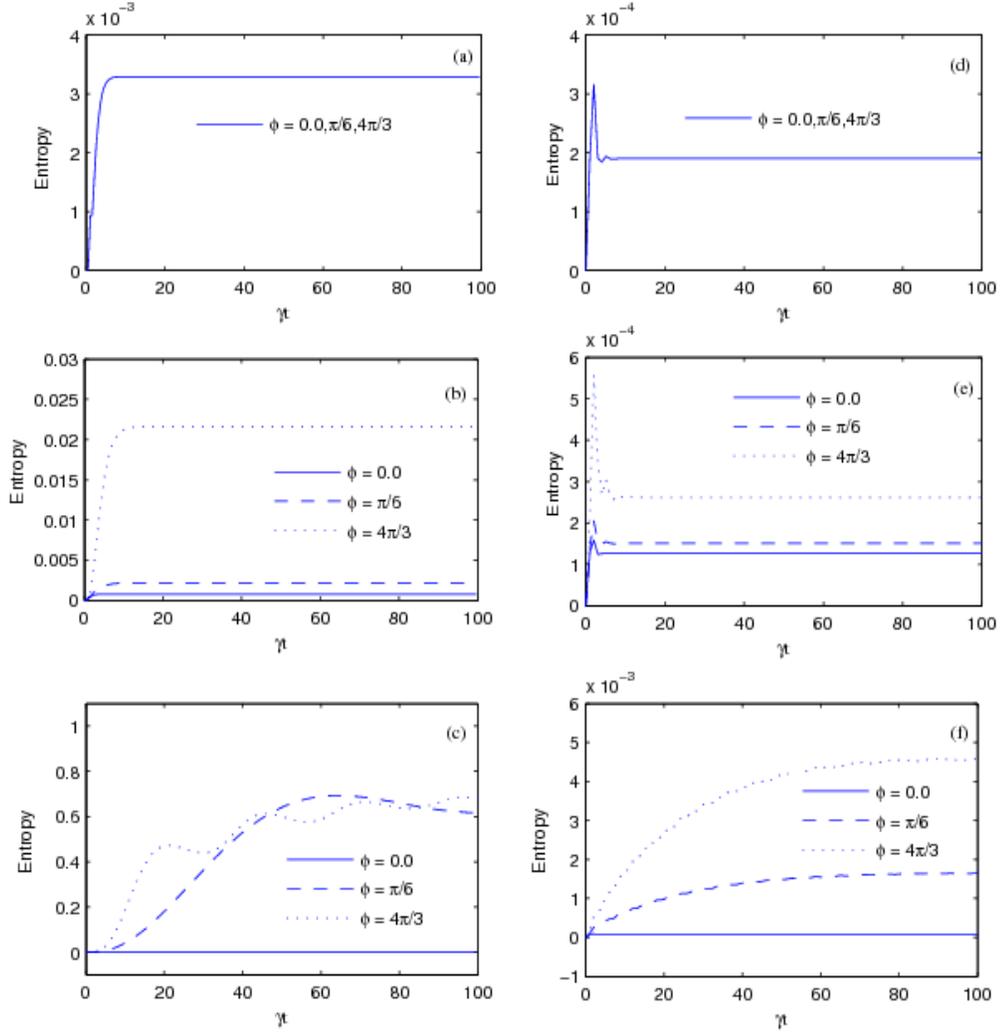

**Figure 3.** The quantum entropy versus detuning. The using parameters are $\gamma = \gamma_{31} = 1$, $\gamma_{21} = 1.0\gamma$, $\Omega_R = \Omega_L = 0.1\gamma$, $\delta = 0.0\gamma$, (a) $K_c = 0$, (b) $K_c = 0.5$, (c) $K_c = 0.99$, for $\varphi = 0$ (Solid), $\varphi = \pi/6$ (Dashed), $\varphi = 4\pi/3$ (Dotted), $\Delta_R = \Delta_L = \Delta$.

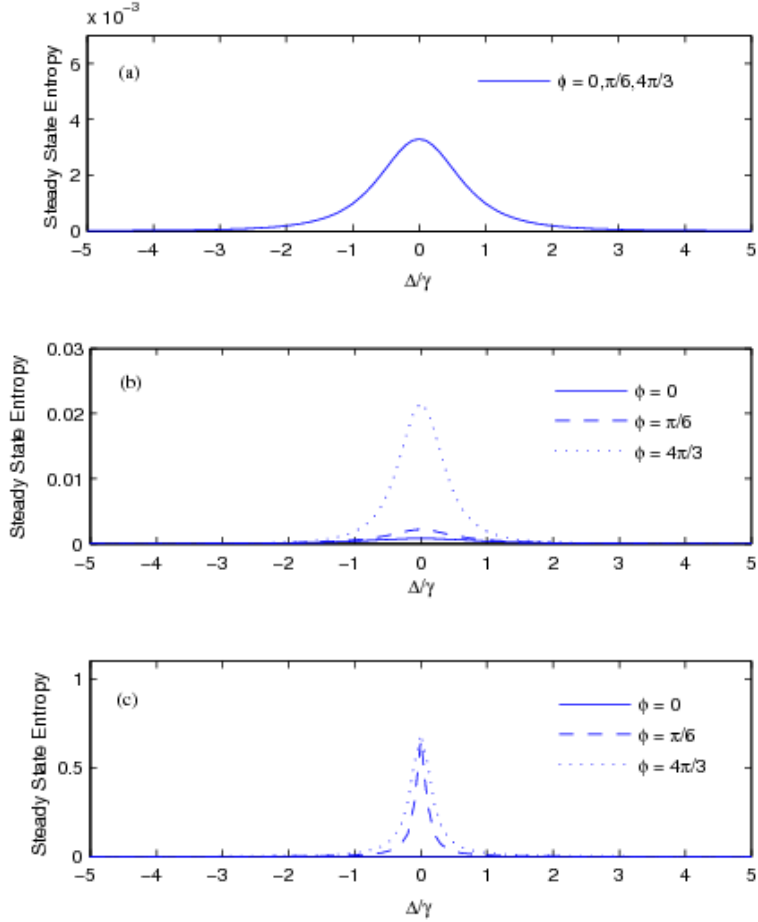

In Figure 4, we show the steady state quantum entropy versus the relative phase of applied fields for two cases of one-photon resonance (a) and beyond it (b). It is clear that in the presence of quantum interference, the steady state entropy changes with respect to the relative phase of applied fields (solid and dashed lines), while in the absence of quantum interference the entanglement is phase-independent (dotted). Moreover, in one-photon resonance condition and for $K_c = 0.99$, $\varphi = 0$, the steady state quantum entropy becomes zero. An investigation on Figure 4(a) and Figure 4(b) shows that, beyond exact one-photon resonance condition, the DEM of the system is negligible.

It is worth to note that in the absence of quantum interference due to the spontaneous emission, (dotted line in Figure 4) the steady state quantum entropy becomes zero for all values of relative phases. Then the quantum interference has a major role in establishing the atom-photon entanglement in a V-type three-level atomic system.

**Figure 4.** The quantum entropy versus the relative phase of applied fields. The parameters are $\gamma_{21} = \gamma_{31} = \gamma = 1.0\gamma$, $\Omega_R = \Omega_L = 0.1\gamma$, $\delta = 0.0\gamma$, $K_c = 0.99$ (solid), 0.5(dashed), 0.0(dash-dotted) for (a) $\Delta_R = \Delta_L = 0.0\gamma$, (b) $\Delta_R = \Delta_L = 2.0\gamma$.

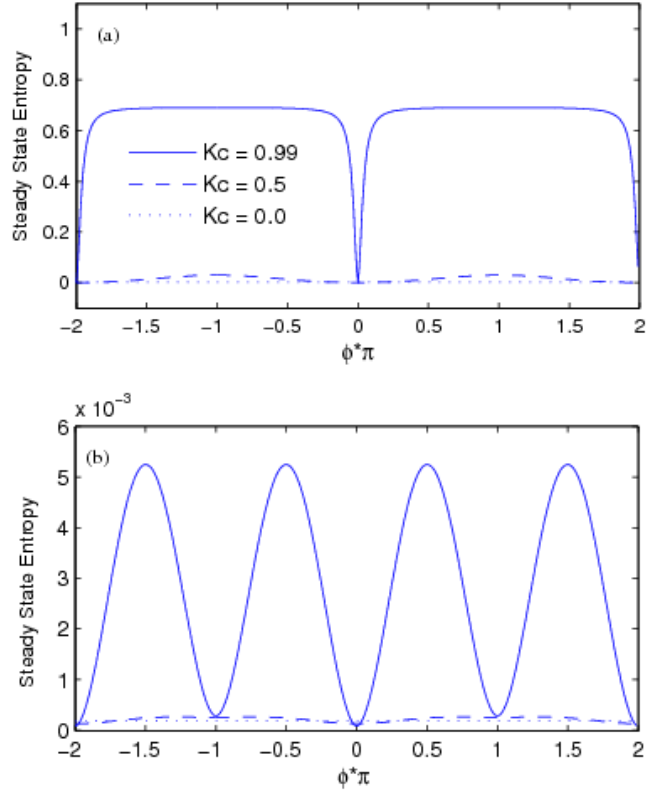

The Rabi frequency of applied fields is another important parameter for controlling the steady state quantum entropy. In Figure 5, we include the effect of quantum interference due to the spontaneous emission and display the steady state quantum entropy versus the relative Rabi frequency $\Omega = \Omega_R / \Omega_L$, for $\varphi = 0$, $K_c = 0.99$, $\Delta_R = \Delta_L = 0.0\gamma$ (solid), $2.0\gamma$ (dashed), $4.0\gamma$ (dotted), and $6.0\gamma$ (dash-dotted). The interesting disentanglement phenomena appears for $K_c = 0.99$, when the relative phase of applied fields is $\varphi = 0$, and the ratio of two Rabi frequencies of applied fields is $\Omega = 1$ [33-35].

**Figure 5.** We display the steady state quantum entropy versus the relative Rabi frequency, $\Omega = \Omega_R/\Omega_L$, for $\varphi = 0$, $K_c = 0.99$, $\Delta_R = \Delta_L = 0.0\gamma$ (solid), $2.0\gamma$ (dashed), $4.0\gamma$ (dotted), $6.0\gamma$ (dash-dotted).

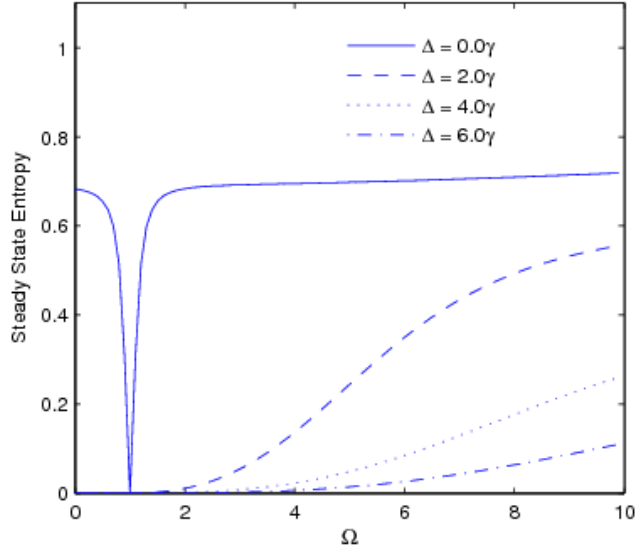

To explain the physical mechanism of such disentanglement, we represent the population behavior of atomic levels versus the relative phase (left column) and the relative Rabi frequencies of applied fields (right column) in Figure 6. The parameters are same as in Figure 5. An investigation on the left column of Figure 6 shows that the disentanglement occurs when all of population is populated in ground state $|1\rangle$. a similar effect appears in the right column. When the one-photon resonance condition is fulfilled (solid lines), just for $\varphi = 0$, all the population remains in the ground state, otherwise the population is distributed in all of three levels of atomic system.

As we have mentioned in the Figure 6, for disentanglement conditions, the population of excited states is negligible and then the steady state entropy becomes zero. We show this point in Figure 7 which is plotted by using the analytical results of Eqs. (3). The analytical solutions are in a good agreement with our numerical results.

**Figure 6.** The population of atomic levels is shown versus the relative phase (left column) and the relative Rabi frequencies (right column) of applied fields. The parameters are $\Delta_R = \Delta_L = 0.0\gamma$ (Solid), $2.0\gamma$ (Dashed), $4.0\gamma$ (dotted), $6.0\gamma$ (Dash-dotted), and $K_c = 0.99$.

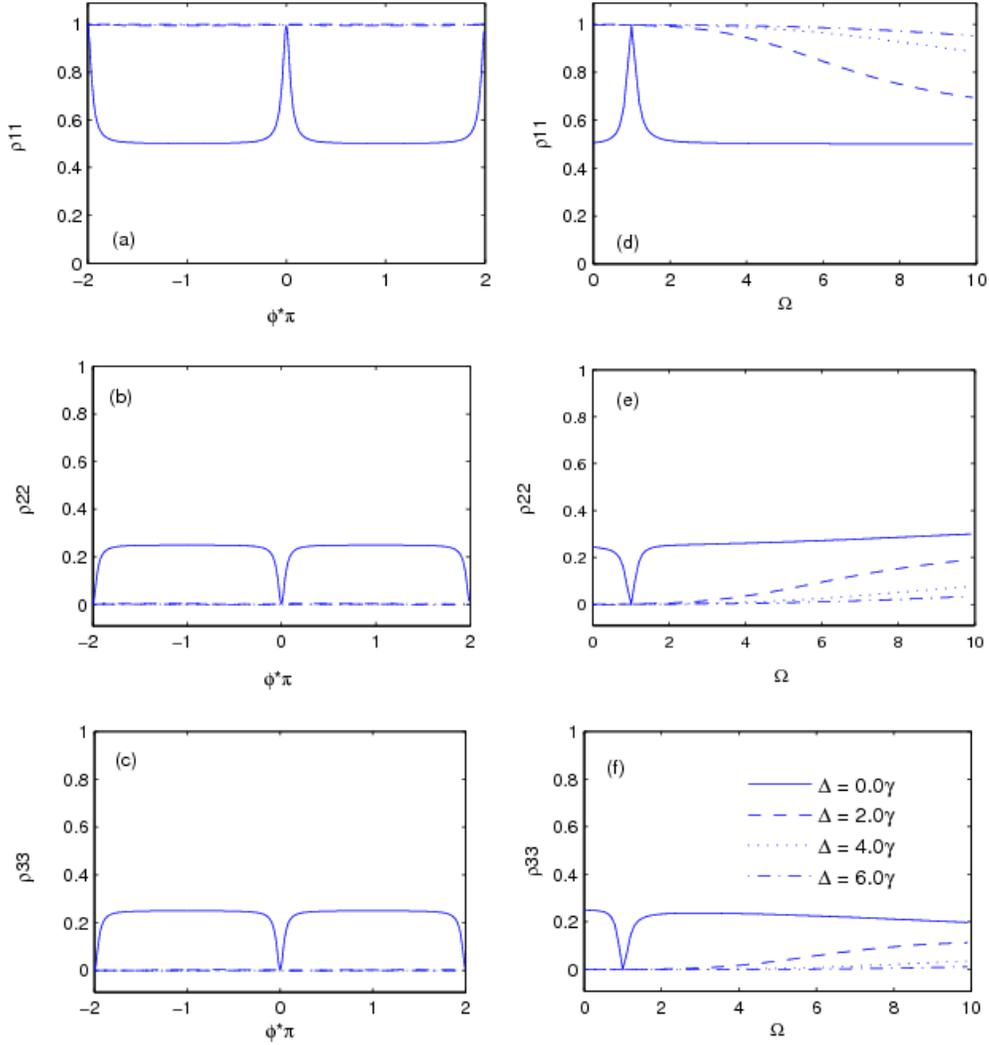

**Figure 7.** The quantum entropy versus relative phase of applied fields for $\gamma = 1$, and relative Rabi frequency $\Omega$ with $K_c = 0.99$, for $\Omega_L = 0.1\gamma$, and $\Delta_R = \Delta_L = 0.0\gamma$.

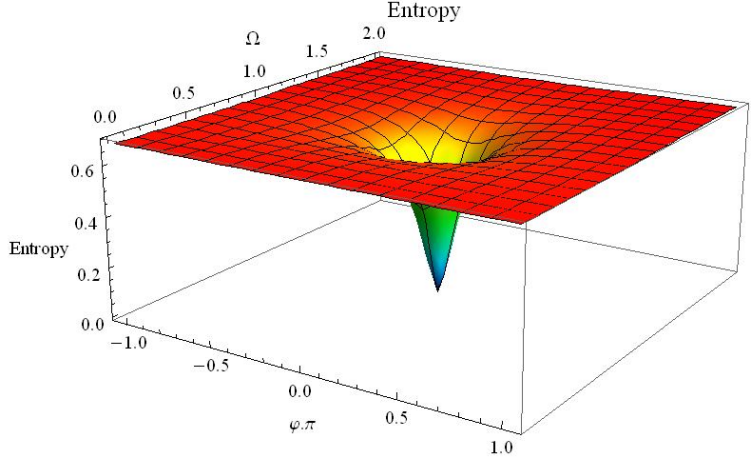

Finally, let us focus on a special and interesting case in which entanglement is non-stationary. In this case, it is shown that one of the eigenvalues of density matrix is zero [37]. Then for calculation of the quantum entropy, according to Eq. (8), it is necessary to note that $\lim_{x \to 0} x \ln x = 0$. To explain the physics of phenomena, let us introduce the new bases $|1\rangle$, $|\psi\rangle = (|2\rangle + |3\rangle)/\sqrt{2}$ and $|\phi\rangle = (|2\rangle - |3\rangle)/\sqrt{2}$. The density matrix Eqs. (2) in these bases can be written as

$$\dot{\rho}_{\psi\psi}(t) = -(\gamma_{21} + \gamma_{31} + 2K_c\sqrt{\gamma_{21}\gamma_{31}}\cos\varphi)\rho_{\psi\psi}(t) - (\frac{\gamma_{21} + \gamma_{31} + i(\Delta_R - \Delta_L)}{2} + iK_c\sqrt{\gamma_{21}\gamma_{31}}\sin\varphi)\rho_{\psi\varphi}(t)$$
$$+ (-\frac{\gamma_{21} + \gamma_{31} - i(\Delta_R - \Delta_L)}{2} + iK_c\sqrt{\gamma_{21}\gamma_{31}}\sin\varphi)\rho_{\varphi\psi}(t) + \frac{i(\Omega_R + \Omega_L)}{\sqrt{2}}(\rho_{1\psi}(t) - \rho_{\psi 1}(t)),$$

$$\dot{\rho}_{\varphi\varphi}(t) = (-\gamma_{21} - \gamma_{31} + 2K_c\sqrt{\gamma_{21}\gamma_{31}}\cos\varphi)\rho_{\varphi\varphi}(t) - (\frac{\gamma_{21} + \gamma_{31} - i(\Delta_R - \Delta_L)}{2} + iK_c\sqrt{\gamma_{21}\gamma_{31}}\sin\varphi)\rho_{\psi\varphi}(t)$$
$$+ (-\frac{\gamma_{21} + \gamma_{31} + i(\Delta_R - \Delta_L)}{2} + iK_c\sqrt{\gamma_{21}\gamma_{31}}\sin\varphi)\rho_{\varphi\psi}(t) + \frac{i(\Omega_R - \Omega_L)}{\sqrt{2}}(\rho_{1\varphi}(t) - \rho_{\varphi 1}(t)),$$

$$\dot{\rho}_{\psi\varphi}(t) = \dot{\rho}^*_{\varphi\psi}(t) = (-\frac{\gamma_{21} - \gamma_{31} + i(\Delta_R - \Delta_L)}{2} + iK_c\sqrt{\gamma_{21}\gamma_{31}}\sin\varphi)\rho_{\psi\psi}(t) + (-\frac{\gamma_{21} - \gamma_{31} - i(\Delta_R - \Delta_L)}{2}$$
$$+ iK_c\sqrt{\gamma_{21}\gamma_{31}}\sin\varphi)\rho_{\varphi\varphi}(t) - (\gamma_{21} + \gamma_{31})\rho_{\psi\varphi}(t) + \frac{i(\Omega_R + \Omega_L)}{\sqrt{2}}\rho_{1\varphi}(t) - \frac{i(\Omega_R - \Omega_L)}{\sqrt{2}}\rho_{\psi 1}(t),$$

$$\dot{\rho}_{1\psi}(t) = \dot{\rho}^*_{\psi 1}(t) = -(\frac{\gamma_{21}+\gamma_{31}+i(\Delta_R+\Delta_L)}{2} + K_c\sqrt{\gamma_{21}\gamma_{31}}\cos\varphi)\rho_{1\psi}(t) - (\frac{\gamma_{21}-\gamma_{31}+i(\Delta_R-\Delta_L)}{2}$$
$$+ iK_c\sqrt{\gamma_{21}\gamma_{31}}\sin\varphi)\rho_{1\varphi}(t) + i(\Omega_R+\Omega_L)(\rho_{\psi\psi}(t)-\rho_{11}(t)) + i(\Omega_R-\Omega_L)\rho_{\varphi\psi}(t),$$

$$\dot{\rho}_{1\varphi}(t) = \dot{\rho}^*_{\varphi 1}(t) = (-\frac{\gamma_{21}-\gamma_{31}+i(\Delta_R-\Delta_L)}{2} + iK_c\sqrt{\gamma_{21}\gamma_{31}}\sin\varphi)\rho_{1\psi}(t) + (-\frac{\gamma_{21}+\gamma_{31}+i(\Delta_R+\Delta_L)}{2}$$
$$+ K_c\sqrt{\gamma_{21}\gamma_{31}}\cos\varphi)\rho_{1\varphi}(t) + i(\Omega_R-\Omega_L)(\rho_{\varphi\varphi}(t)-\rho_{11}(t)) + i(\Omega_R+\Omega_L)\rho_{\psi\varphi}(t),$$

$$\dot{\rho}_{11}(t) = -(\dot{\rho}_{\varphi\varphi}(t)+\dot{\rho}_{\psi\psi}(t)).$$

level. For our interesting case $K_c = 1.0$, $\varphi = \pi$, $\Omega_R = \Omega_L = \Omega_0$ and $\rho_{11}(0)=1$, Eqs. (9) converts to the simple following expressions,

$$\dot{\rho}_{\phi\phi}(t) = 0,$$

$$\dot{\rho}_{\psi\psi}(t) = \frac{2i\Omega_0}{\sqrt{2}}(\rho_{1\psi}(t)-\rho_{\psi 1}(t)),$$

$$\dot{\rho}_{11}(t) = -\dot{\rho}_{\psi\psi}(t),$$

$$\dot{\rho}_{1\psi}(t) = \dot{\rho}^*_{\psi 1}(t) = 2i\Omega_0(\rho_{\psi\psi}(t)-\rho_{11}(t)), \qquad (10)$$

$$\dot{\rho}_{\psi\phi}(t) = \dot{\rho}^*_{\phi\psi}(t) = 0,$$

$$\dot{\rho}_{1\phi}(t) = \dot{\rho}^*_{\phi 1}(t) = 0,$$

The analytical solutions of Eqs.(10) are given by $\dot{\rho}_{\phi\phi}(t) = 0$,

$$\rho_{\psi\psi}(t) = \frac{1}{2}(1-\cos(\frac{4\Omega_0}{\sqrt[4]{2}}t)),$$

$$\rho_{11}(t) = \frac{1}{2}(1+\cos(\frac{4\Omega_0}{\sqrt[4]{2}}t)), \qquad (11)$$

$$\rho_{1\psi}(t) = \rho^*_{\psi 1}(t) = \frac{-i\sqrt[4]{2}}{2}(\sin(\frac{4\Omega_0}{\sqrt[4]{2}}t)),$$

$$\rho_{1\phi}(t) = \rho_{\phi 1}(t) = \rho_{\psi\phi}(t) = \rho_{\phi\psi}(t) = \rho_{\phi\phi}(t) = 0.$$

These equations imply that the proposed three-level scheme reduce to a two-level quantum system with simple oscillatory behavior. The population of levels $|1\rangle$ and $|\psi\rangle$ show the oscillatory behavior, however the level $|\varphi\rangle$ is decoupled and then the corresponding population is zero and the density matrix is converted to a $2\times 2$ matrix. The corresponding eigenvalues are given by

$$\lambda_\pm = \frac{1 \pm \sqrt{\cos^2(\frac{4\Omega_0}{\sqrt[4]{2}}t) + \sqrt{2}\sin^2(\frac{4\Omega_0}{\sqrt[4]{2}}t)}}{2}. \qquad (12)$$

Then such oscillatory eigenvalues apply an oscillatory dynamical behavior to the quantum entropy. In this special case no stationary solution can be found for the entanglement of the system.

**5. Conclusion and Perspectives**

We have investigated the effect of quantum interference due to the spontaneous emission on the dynamical behavior of atom-photon entanglement in a V-type three-level quantum system by using the atomic reduced entropy. It is shown that in the presence of quantum interference of spontaneous emission the entanglement of the atom-photon can be controlled either in intensity or by the relative phase of applied fields. Moreover, it is demonstrated that for the special parameters, disentanglement occur in this system.

**Acknowledgements**

We would like to thanks Prof. Girish. S. Agarwal for helpful discussions and comments on this work.